\documentclass[prd,a4paper,nofootinbib,
showpacs, 
preprint
]{revtex4}
\usepackage{graphicx}
\def\eq#1{{eq.~(\ref{#1})}}
\def\eqs#1#2{{eqs.~(\ref{#1})--(\ref{#2})}}

\def\vev#1{\left\langle #1\right\rangle}

\def\Im{\mbox{Im}\,}
\def\Re{\mbox{Re}\,}
\def\Tr{\mbox{Tr}\,}

\def\hbar{\hspace{0pt}\raisebox{1pt}{$-$} \hspace{-7pt} h}

\def\5{\overline 5}

\newcommand{\be}{\begin{equation}}
\newcommand{\ee}{\end{equation}}
\newcommand{\bea}{\begin{eqnarray}}
\newcommand{\eea}{\end{eqnarray}}
\newcommand{\nn}{\nonumber}

\begin{document}
\title[Littlest Higgs]{The large mass of the littlest Higgs 
 }
\date{\today
}
\author{F.~Bazzocchi}
\author{M.~Fabbrichesi}
\affiliation{INFN, Sezione di Trieste and\\
Scuola Internazionale Superiore di Studi Avanzati\\
via Beirut 4, I-34014 Trieste, Italy}
\author{M.~Piai}
\affiliation{Department of Physics, Sloane Physics Laboratory\\
University of Yale, 217 Prospect Street\\
New Haven CT 06520-8120, USA}
\begin{abstract}

\noindent  We study the exact (one-loop) effective potential of the littlest Higgs model
and determine the dependence of physical quantities, such as the vacuum
expectation value $v_W$ and mass $m_h$ of the Higgs boson, 
on the fundamental parameters
of the Lagrangian---masses, couplings of new states, the fundamental 
scale $f$ of the sigma model, and the  coefficients of operators quadratically sensitive to the cutoff of the theory.
On the one hand, we show that it is possible to have the electroweak ground state and  a relatively large 
cutoff $\Lambda = 4\pi f$ with $f$  in the 2 TeV range without requiring unnaturally small 
coefficients for quadratically divergent quantities, and with only moderate 
cancellations between the contribution of different sectors to the 
effective potential of the Higgs. 
On the other hand, this cannot be achieved while at the same
time keeping $m_h$ close to its  current lower bound of $114.4$ GeV.
The natural expectation for $m_h$ is $O(f)$, mainly because of  large logarithmically  divergent contributions to the effective potential of the top-quark sector.
Even a  fine-tuning at the level of $O(10^{-2})$ in the coefficients of the quadratic divergences
is not enough to produce small physical Higgs masses, and the natural expectation is in
the 800 GeV range  for $f \sim 2$ TeV.
We conclude that the littlest Higgs model is a solution of the 
little hierarchy problem, in the sense that it stabilizes the electroweak symmetry breaking 
scale to be a factor of 100 less than the cutoff of the theory, but this
requires a quite large physical mass for the Higgs, and hence precision electroweak
studies should be redone accordingly. 
We also study finite temperature corrections.
The first order electroweak phase transition is no stronger
than in the standard model. A second phase transition (non-restoration of symmetry at high temperature)
depends strongly on the logarithmic terms in the potential.
\end{abstract}
\pacs{12.60.Fr, 11.30.Qc}
\maketitle
%
\vskip1.5em
\section{Motivations} 
\label{sec:mb}
 
The littlest Higgs model~\cite{littlest}  has been introduced  to address the little hierarchy problem which arises
because of the quadratically divergent one-loop corrections to the quadratic term $\mu^2_h h^2$ in the standard model Higgs boson potential. Given a cutoff of the theory at $\Lambda$, these corrections are of order $\Lambda^2/16\pi^2$.
For a Higgs boson  mass $m_h$ around or just above the current bound of 114.4 GeV~\cite{PDG},  the cutoff must be around 1 TeV to be natural. This is in (mild) contrast with respect to  current bounds on new physics coming from electroweak precision measurements~\cite{PDG}  which suggest the absence of new physics up to  10 TeV. The same problem arises if we consider  the electroweak vacuum $v_W$ and wish to stabilize its value in the presence of a 10 TeV cutoff.

In the littlest Higgs model (and in similar models built according to the same idea~\cite{models,models2,littleflavons}) this problem is solved by  making the Higgs boson into a pseudo-Goldstone mode of a global symmetry  $SU(5)$ (containing two copies of the electroweak groups $SU(2)\times U(1)$) spontaneously broken at the scale $f$. The model is thus defined up to a cutoff $\Lambda=4\pi f$. The scalar  sector of the theory   is protected by two copies of a global $SU(3)$ symmetry which are only  broken collectively by two or more terms in the lagrangian so that at one-loop the Higgs boson mass only receives logarithmically divergent contributions and $v_W$ is much less than $f$. 

While it is clear that this idea works qualitatively, there are two tests that the model must  pass  to be also quantitatively successful. First of all, because of the enlarged symmetry, the littlest Higgs model contains more states (heavy gauge bosons, a scalar $SU(2)$ triplet and at least one heavy quark) than the standard model and their effect  on electroweak precision measurements constrains the possible values of the symmetry breaking scale $f$.  For the model to work, these constraints must not be too strong and therefore still allow a value of $f$ around 2 TeV. This seems to be the case in the most recent fit in which loop corrections and the effect of the scalar triplet  are properly included~\cite{CD} (for previous, and more pessimistic, analyses~see \cite{precision}). 

The second test has to do with the fine-tuning in the Coleman-Weinberg (CW) effective potential~\cite{CW} for the Higgs boson (and the closely related isospin triplet). 
The CW potential  contains both quadratically  and logarithmically divergent terms. These  divergent terms are controlled by (unknown) coefficients, the determination of which would require the ultraviolet completion of the theory~\cite{TC}. For all practical purposes, they can be considered additional parameters of the model, and of them, only those in the quadratically divergent terms are relevant since the others can only weakly affect the potential. The meaning of these coefficients is the amount of symmetry breaking we allow into the model from  operators induced by states living below (which are known) and just above  the cutoff (which are instead unknown). For the model to be natural, these coefficients cannot be very small because this would be equivalent to suppressing by hand the symmetry breaking operators.

To address the question of how much more natural is the littlest Higgs model  with respect to the standard model, we study the  exact potential, rather than its truncation to terms quartic in the fields,  and include all logarithmic terms---which are usually neglected in all analysis~\cite{precision,CD}.  These logarithmic terms are important and cannot be neglected; for the model to be successful they must be numerically small enough to give the Higgs boson the desired mass without further fine tuning. 

In reporting on our results, we first illustrate the main results with an approximate analysis,
in which we neglect the existence of the triplet field, and expand the resulting
potential for the Higgs field in powers of $h/f$. The numerical study
however is performed using the complete 1-loop potential,  with inclusion
of the triplet field, and without expanding in $h/f$ in looking for the
minima  for the vacuum expectation values of the scalar doublet and triplet.
 
After  fixing three combinations of gauge and Yukawa couplings to reproduce the standard model couplings $g$, $g'$ and $\lambda_t$ (that is, the mass $m_t$ of the top quark), the littlest Higgs model is controlled by six parameters: 2 gauge and 1 Yukawa couplings, the two coefficients $c_1$ and $c_2$, of the quadratically divergent terms, one for the bosonic and one for the fermionic loops and the symmetry breaking scale $f$. At the same time, we have six constraints given by the vanishing of the first derivatives in the doublet and triplet directions, the value $v_W$ of the electroweak vacuum (that is, the value of the Higgs field in the minimum of the potential) and that of the triplet field, the mass $m_h$ of the Higgs boson and  of the triplet (the second derivatives of the potential at the minimum).

We therefore have an effective theory in which all parameters and coefficients are constrained and the  model completely determined. 
We can study it as a function of the physically significant parameters $v_W$, $m_h$ and $f$; in particular, what are the values of the coefficients $c_1$ and $c_2$? Are there any choices which allow for $v_W$ at its physical value, the mass of the Higgs to be, say, around 115 GeV  and $f$  around  2 TeV, as suggested by the electroweak data? While the answer seems to be positive for the value of  $v_W$ (and in this respect the model is successful), it is negative for $m_h$, in the sense that there are no solutions, as we vary the gauge couplings and the coefficients, leading to $m_h$ and $f$ in the desired range. The main reason for this failure lays in the logarithmic contributions to the Higgs boson mass which are $O(f)$ rather than $O(m_h)$ thus leading to a  littlest Higgs with a mass  around 800 GeV. Larger masses are also possible (and natural) but would lead the theory outside its  perturbative definition.

At first, this negative result does not seem too troublesome since we know that the inclusion of the next-order (two-loop) corrections is crucial in the precise determination of $m_h$.  What is surprising is the large size of these logarithmic terms which force us to introduce a proportionally large two-loop correction thus rising some doubts on the entire perturbative expansion. Even after the two-loop corrections have been included, the possible choices in which the model gives $m_h$ and $f$ in the  desired range lead to very unnatural values of the coefficients---at least one of the coefficients $c_i$ must be unreasonably small---which  defy the very purpose of introducing the model. In fact, as we already pointed out, these coefficients control the symmetry breaking operators, and if we were allowed to  suppress them by fine tuning we could have done it directly in the standard model without having to resort to the littlest Higgs model in the first place.

This problem  seems to be more serious for the model than  the amount of fine-tuning in the parameters imposed by electroweak precision measurements. Moreover, our analysis shows that the recent fit within the littlest Higgs model of the electroweak radiative corrections~\cite{CD} falls in a region of the parameter space that is excluded by the requirement of having the ground state near zero rather than $f \pi/2$.

We consider next an improved version of the littlest Higgs model in which the top fermionic sector  is completed to make its contribution to the CW one-loop potential  finite~\cite{TC}. We study this model and show that, even though (marginally) better than the littlest Higgs model, again the requirement of a  light Higgs boson mass  and $f$ around 2 TeV would lead to unreasonable values of the parameters and excessive fine tuning.

Having computed the exact potential, it is interesting to also study  physics at finite temperature $T$. There are two issues. The first is about the electroweak transition and whether is of the first order and any stronger than in the standard model for values of the Higgs mass around the current bound. This question has a negative answer. The second is about the symmetry non-restoration which many models based on pseudo Goldstone bosons present. The details of the high-$T$ phase transition is very sensitive to the values of  the coefficients of the divergent terms. We show that it is also sensitive to the logarithmically divergent terms which therefore cannot be neglected.


\vskip1.5em
\section{The exact potential} 
\label{sec:tep}

The littlest Higgs model is based on an  approximate $SU(5)$ global symmetry spontaneously broken to $SO(5)$.  The symmetry breaking  gives rise to $14$  Goldstone bosons. Four of them are eaten by  the heavy gauge bosons that acquire a mass when $[SU(2)\times U(1)]^2$ is broken to the diagonal  $[SU(2)\times U(1)]$, which is then  identified with the electroweak gauge group. The other $10$ degrees of freedom with respect to the  diagonal  $[SU(2)\times U(1)]_W$ give rise to two complex fields: a $SU(2)$ triplet $\phi$ and doublet $\varphi$. 

When the  $[SU(2)\times U(1)]_W$ gauge group is broken to the $U(1)_Q$ electric charge gauge group, other $3$ degrees of freedom are eaten by the standard model gauge bosons, and the remaining physical fields are a double charged complex scalar $\phi^{++}$, a single charged  complex scalar $\phi^{+}$, one neutral pseudoscalar,  $\phi^{0}$, and two neutral scalars, $t$ and $h$, the Higgs boson. The two neutral scalars arise from the mixing between the neutral components of $\Im{\phi}$ and $\Re{\varphi}$, which are the components that acquire a vacuum expectation value and break the electroweak gauge group into $U(1)_Q$. 

In the following we will consider only the part of the effective potential involving the scalar components responsible of the  electroweak gauge group spontaneous breaking. We will come back to the complete spectrum when we will discuss the littlest Higgs at  finite temperature. 

 The lagrangian for the Goldstone bosons $\Sigma$ is given by
\bea
\mathcal{L}_\Sigma&=& \mathcal{L}_K + \mathcal{L}_t\,,
\eea
where $ \mathcal{L}_K $ is the kinetic term 
\be
\mathcal{L}_K= \frac{f^2}{8}\Tr{\big(D_\mu \Sigma \big)\big(D^\mu \Sigma^* \big)}\,,
\ee
$f$ is the $SU(5)$ spontaneous breaking scale and 
\be
D_\mu= \partial_\mu  -\sum_i \{i g_i W_{i_{\mu}}^a (Q_i^a \Sigma +\Sigma Q_i^{a\,T}) + i {g'}_iB_i(Y_i\Sigma +\Sigma Y_i^T)\}\,,
\ee
where $g_i$ and ${g'}_i$ , $i=1,2$, are the gauge couplings of the two copies of $[SU(2)\times U(1)]_i$ gauge groups. 

$\mathcal{L}_t$ is the top quark Yukawa lagrangian 
\be
\mathcal{L}_t= \sqrt{2} \lambda_1 \,f\,\epsilon_{ijk} \epsilon_{xy}\, \chi_i\, \Sigma_{jx}\,\Sigma_{ky} {u'}_3^c + \sqrt{2} \lambda_2 \,f\,\tilde{t} \tilde{t^c} + h.c. \, ,
\ee
where $\chi_i$ is a triplet of one of the two  $SU(3)$ groups in $SU(5)$ and $\tilde t$ a vector-like quark~\cite{littlest}.
 
The effective potential of the Higgs boson in the littlest Higgs model is found by computing the CW potential~\cite{CW} generated by the gauge boson and fermion loops.  At the one-loop, it can be written as
\bea
\label{V0}
V_1 [c_i,g_i,g_i',\lambda_i;\,  \Sigma]  & = &  3\, \frac{ c_1 \Lambda^2 }{32 \pi^2} \Tr M^2_B(\Sigma) -
12 \, \frac{ c_2 \Lambda^2 }{32 \pi^2} \Tr M^2_F(\Sigma) + 3\, \frac{1}{64 \pi^2} \Tr M_B^4(\Sigma) \log c_3 M^2_B(\Sigma)/ \Lambda^2 \nn \\
& &  -12\, \frac{1}{64 \pi^2} \Tr M_F^4(\Sigma) \log c_4 M^2_F(\Sigma)/ \Lambda^2  \label{pot} \, ,
\eea 
where the factors 3 and 12 in front of the operators count the degrees of freedom of, respectively, bosons and colored fermions.
The coefficients $c_i$ are unknown constants, the values of which come (presumably) from the ultraviolet completion of the theory~\cite{TC}. They are there because these terms are divergent and UV physics cannot be safely  decoupled. Additional states may contribute to the relevant operators and their effect cannot be computed.
From the effective theory point of view, these coefficients are arbitrary numbers to be determined. In what follows,  we take $c_{3,4}$ equal to 1 since they appear in the logarithmic contributions and their contribution cannot be crucial~\footnote{The presence of a divergence
signals the necessity to add  a counterterm in the theory, and hence, as for quadratic divergences,
one should allow for the coefficient of this term to vary. On the other hand, 
the divergent part is local in $\Sigma$, while we find that the most significant 
contribution to the potential comes from the  (finite) non-local  part, which does not depend on the arbitrary coefficients $c_i$.
This fact will be exemplified in
a cleaner  way when we discuss the modification of
the top sector which removes 1-loop divergences completely.
Modifying the coefficients $c_3$ and $c_4$ certainly affects  the potential,
but does not change significantly our results, 
unless extremely big or extremely small choices are made,
which would imply very big fine-tuning.}.

As we shall see, it is also important to include   terms that may arise from two-loop quadratic divergent contributions.  They can be of various (and complicated) forms, and we indicate it by a generic operator of canonical dimension two:
\be
V_2[c_5;\, \Sigma] =  \frac{c_5 \Lambda^2}{(4 \pi)^4} {\cal O}_{2-loop}(\Sigma) \, .
\ee
We are not going to compute these terms and, as discussed below,  just take $c_{5}$  to be the coefficient of  a term of order $f^2/16 \pi^2$, which controls the size of the two-loop quadratically divergent contributions.

   The traces in \eq{pot} over the effective (squared) masses are the one-loop quadratically divergent contribution of, respectively, bosonic and fermionic degrees of freedom in the CW potential:
   \bea
\label{massef}
   \Tr M^2_B(\Sigma) & = & \frac{f^2}{4}\Big( g_i^2  \, \sum_a \Tr [(Q^q_i \Sigma)(Q^q_i \Sigma)^*]+ {g'}_i \Tr{ [(Y_i \Sigma)(Y_i \Sigma)^*]}\Big) \nn \\
   \Tr M^2_F(\Sigma) & = & - 2 \, \lambda_1^2\, f^2 \epsilon^{wx}\epsilon_{yz} \epsilon^{ijk}\epsilon_{kmn} \Sigma_{iw}\Sigma_{jx}\Sigma^{*my}\Sigma^{*nz}
 \eea  
 These terms depend on the model coupling constants $g_i$, ${g'}_i$ and $\lambda_{1,2}$. The gauge couplings  can be rewritten as functions of the $SU(2)$ and $U(1)$ electroweak   $g, {g'}$ gauge couplings and of two new parameters $G, {G'}$ defined by
\bea
G^2 &=& g_1^2+g_2^2 \nn\\
{G'}^2 &=& {g'}_1^2+ {g'}_2^2 \,.
\eea
Since 
\bea
g^2 = \frac{g_1^2 g_2^2}{g_1^2+g_2^2} &\quad  & {g'}^2= \frac{{g'}_1^2 {g'}_2^2}{{g'}_1^2+ {g'}_2^2 } \,,
\eea
we have
\be
\label{ridefg}
g^2_{1,2} = \frac{G^2}{2} \pm \frac{G}{2}\sqrt{G^2- 4 g^2} \,,
\ee
 and similar  expressions for  the $U(1)$ ${g'}_i$ couplings. The  standard model gauge couplings are given by $g=\sqrt{4 \pi \alpha/\sin \theta_W}$ and $g' = g \tan \theta_W$ in terms of the fine structure constant $\alpha$ and the Weinberg angle $\theta_W$.

In a similar way, by imposing that  the top-quark Yukawa coupling $\lambda_t$ gives the experimental mass $m_t$, $\lambda_{2}$ can be expressed in terms of  $\lambda_t$ and  $\lambda_1$ which we  rename $x_L$. From
\bea
\label{ridefyuk}
\lambda_t &=&\frac{2 \lambda_1 \lambda_2}{\sqrt{\lambda_1^2+\lambda_2^2}} \nn\\
x_L&=&\lambda_1 \,,
\eea
we have
\be
\lambda_2= \frac{x_L \lambda_t}{\sqrt{4 x_L^2 -\lambda_t^2}} \, .
\ee

Together, \eqs{ridefg}{ridefyuk} fix the range of the parameter $G$,${G'}$ and $x_L$. By imposing the reality of $g_i$, ${g'}_i$ and $\lambda_t$ we have
\bea
G \,\geq\, 2 g(m_W)  & \quad
{G'}\, \geq \, 2 {g'}(m_W) &\quad
x_L \,\geq  \,\frac{\lambda_t(m_W)}{2} \, .
\eea

The value $G= 2 g$ corresponds to the maximally symmetrical case where  $g_1=g_2$ and the heavy bosons are decoupled from their lighter copies. The actual value is usually chosen so as to minimize the overall electroweak corrections~\cite{precision,CD}.

By combining \eq{massef} with \eqs{ridefg}{ridefyuk}, the part of the CW potential proportional only to the Higgs boson  field $h$ is
\bea
\Tr M^2_B &=& \frac{3 f^2}{4}  G^2 +  \frac{f^2}{20} {G'}^2 +  \frac{f^2}{16} (G^2 + {G'}^2) \sin^4 h/f  \nn \\
\Tr M^2_F & = &  \frac{8 f^2  x_L^4}{4 x_L^2 -\lambda_t^2}  - 2 f^2 x_L^2 \sin ^4 h/f  \,,\eea
where $h= \Re{\varphi^0}$, with $\varphi^0$ the neutral components of the complex $SU(2)$ doublet $\varphi$. Notice the cancellation of terms proportional to $\sin ^2 h$.

The complete expression inclusive of the triplet field is too complicated to be reported here.  We only write the contribution to the triplet mass
\be
 \frac{3 f^2}{4} \left[ c_1 (G^2 + G'^2) + 64 c_2 x_L^2 \right] 
\ee
because it will be important in discussing the relevance of the logarithmic corrections.

Once these coupling constants have been fixed together with the coefficients $c_i$'s, the model is completely determined and, after requiring  the potential (\ref{pot}) to have  a minimum in $\vev{h} =v_W/\sqrt{2}$, we can study the  values of $f$ and $m_h$ which are possible within the littlest Higgs model. Vice versa, by imposing the desired values for $f$ and $m_h$, we can find what are the required values for these coefficients.

The terms in (\ref{pot}) proportional to logarithms of the cutoff give rise to the Higgs boson mass but also contribute to the other terms in the potential. The latters are usually neglected~\cite{precision,CD}. As it turns out, they are important and, as we shall show, crucial in determining the properties of the model. Their main contribution is to the quadratic terms of the potential which we write as 
\be
\label{mulog}
 \mathcal{L}_{log}= \mu_h^2 h^2 + \mu_{t}^2 t^2 \,, \label{mu2}
\ee
where  $h$ and $t$ are defined as  $\vev{\Re{\varphi^0}}$ and  $\vev{\Im{\phi^0}}$ respectively and where we have neglected the  subleading term $\mu_{th} h t h$.

Taking only the leading order of each term of \eq{mulog}, we have
\bea
\label{esplog}
\frac{\mu_h^2}{f^2} &=& - \frac{9}{256 \pi^2} g^2 G^2 
\log \frac{G^2}{64 \pi^2} - 
 \frac{3}{1280 \pi^2} q^2 G'^2 
 \log \frac{G'^2}{320 \pi^2} \nn \\
&+&\frac{3}{\pi^2}\,
\frac{\lambda_t^2 x_L^4}{4 x_L^2-\lambda_t^2}\,
\log  \frac{4 x_L^4}{8 \pi^2 (4 x_L^2-\lambda_t^2)}   \nn \\
\frac{\mu_t^2}{f^2}  &=& \frac{3}{128 \pi^2}\left[ \frac{G^2}{2}(G^2- 8 g^2) \log  \frac{G^2}{64 \pi^2}  + \frac{G'^2}{10}(G'^2- 4 g'^2)\log \frac{G'^2}{320 \pi^2} \right]  \nn\\
 &+&   \frac{24}{\pi^2}\, \frac{x_L^6}{4 x_L^2- \lambda_t^2}\,\log \frac{4 x_L^4}{8 \pi^2 (4 x_L^2- \lambda_t^2)}  \, .
\eea

We have written \eq{esplog}  as function of the electroweak parameters $g$, ${g'}$ and $\lambda_t$, and of the free parameters, $G$, ${G'}$ and $x_L$, only for convenience and used $\Lambda= 4 \pi f$. Usually~\cite{precision,CD},  the terms in the potential are reported as functions of the heavy gauge bosons and of the heavy top-like quark  masses, which by \eqs{ridefg}{ridefyuk} are given by
\be
M^2_{W'} = \frac{1}{4}G^2 f^2 \quad  M^2_{B'}=\frac{1}{20} G'^2 f^2 \quad  M^2_{\tilde{t}}= \frac{8 x_L^4}{4 x_L^2- \lambda_t^2} f^2 \, .
\ee 

\subsection{Approximate analysis}
\label{apan}
Before embarking in the   analysis of the complete model, it is 
useful to examine   qualitatively  its main features. This will help 
in elucidating the numerical analysis in the next section. In particular, there are two conditions we would like to satisfy: for  $f\simeq$ 2 TeV and $v_W$ at the electroweak scale we must have
\be
\frac{v_W^2}{f^2} = -\frac{2 \mu^2_h}{\lambda f^2} \simeq 10^{-2},
\ee
while, at the same time, in order to have a light $m_h$
\be
\frac{\mu^2_h}{f^2} \simeq 10^{-3}\, .
\ee

For 
simplicity,  we  ignore the $U(1)$ groups and therefore take  
${g'}_{1,2}=0$.
We expand \eq{V0} up to the fourth and second order in the  doublet 
and triplet field components respectively, so that the potential is 
given by
\be
\label{V0ap}
V[h,t]=\mu_h^2  h^2+ \lambda_3 \, h t h + \lambda_4 h^4 + 
\lambda_{\phi} t^2 \, .
\ee
 If in \eq{V0ap} we neglect the logarithmic contributions (except in 
$\mu_h^2$), the coefficients $\lambda_3$, $\lambda_4$ and 
$\lambda_{\phi}$ are easily obtained from \eq{massef} and they are 
given by 
\bea
\label{espcomp}
\lambda_{\phi}/4 & = \lambda_4 =& \frac{3}{16}( c_1 G^2 + 64 c_2  
x_L^2) \nn \\
\lambda_3 &=& \frac{3}{4} ( c_1 G^2(s^2_g-c^2_g)  + 64 c_2 x_L^2) \,,
\eea
where $G$ and $x_L$ have been defined in the previous section and 
\bea
c_g= g_1/G &\quad & s_g = g_2/G  \,.
\eea

By imposing the conditions for the existence of a minimum in the 
potential,
\bea
\frac{\partial V[h,t]}{\partial h} &=& 0\, , \nn \\
\frac{\partial V[h,t]}{\partial t} &=& 0 \, , 
\eea
we find that the vacuum expectation values are given by 
\bea
\vev{t}&=& -\frac{\lambda_3}{2 \lambda_{\phi}} \frac{\vev{h}^2}{f} 
\nn\\
\vev{h}^2&=& -\frac{\mu_h^2}{\tilde{\lambda}} \,,
\eea
where 
\bea
\label{leff}
\tilde{\lambda}& = &2 \lambda_4 - \frac{\lambda_3^2}{2 
\lambda_{\phi}} \nn \\
&=&\frac{3}{2}  \frac{c_1 G^2 c_g^2 ( c_1 G^2 s_g^2 + 64 c_2 x_L^2)}{  c_1 G^2  + 64
c_2 x_L^2} \,. \label{aa}
\eea

Assuming $c_1=c_2=1$, and hence no fine-tuning between UV and low 
energy
sources of symmetry breaking, this reduces to:
\be
\tilde{\lambda} = \frac{3}{2} \frac{G^2 c_g^2 ( G^2 s_g^2 + 64 x_L^2)}{G^2  + 64  x_L^2} \,.
\ee

In order to make contact with electroweak physics we have to impose 
$\vev{h}=v_w/\sqrt{2}$, where $v_W$ is 246 GeV.
The mass of the physical Higgs boson, $H$, and of the physical 
scalar are therefore
\bea
\label{manalit}
m^2_h = -2 \mu_h^2 = \tilde{\lambda} v_w^2 \, ,\nn\\
m^2_{\phi}= \lambda_{\phi} f^2 \,.
\eea

Finally, let us give  an estimate of $\mu_h^2$, which  is determined 
by  the logarithmically 
divergent part of the CW potential plus finite terms.
At leading order, \eq{esplog} yields
\bea
\label{mHlog}
\frac{\mu_h^2}{f^2} &=& - \frac{9}{256 \pi^2} g^2 G^2 
\log \frac{G^2}{64 \pi^2}  +\frac{3}{\pi^2}
\frac{\lambda_t^2 x_L^4}{4 x_L^2-\lambda_t^2}
\log  \frac{4 x_L^4}{8 \pi^2 (4 x_L^2-\lambda_t^2)}  
\,,
\eea
and using the constraints, on $m_t$ and the gauge couplings,
\bea
x_L&>&\lambda_t/2\,\nn\\
G^2&>&4g^2\,,
\eea
we find  that, for instance by taking $x_L\simeq 1$ and $G\simeq 2 
g$, we have 
\bea
\label{mHlognum}
\mu_h^2&\simeq & \left( 0.01 g^2 G^2 - \frac{x_L^4}{4 x_L^2-1} 
\right) \,f^2 \simeq -0.3 f^2\,.
\eea
With these, one gets for the Higgs mass:
\be
m^2_h \,\simeq\,0.6 f^2\,.
\ee 
If we want $f \simeq 2$ TeV, then 
\be
\tilde{\lambda}  =  -\frac{2\mu_h^2}{v^2_{w}}=\frac{0.6 f^2}{v^2_{w}}\,\simeq\,40\,,\
\ee
which is at the limit of validity of the perturbative expansion (the 
expansion parameter being
roughly given by $\tilde{\lambda}/16\pi^2$).
The mass of the triplet would be $m_{\phi} \simeq$ few TeV.
This scenario would correspond to a cutoff of the theory $\Lambda 
\simeq 25$ TeV,
which is what we wanted, 
but requires a mass for the physical Higgs $m_h \simeq 1$ TeV.

On the contrary, if we also demand that the Higgs boson mass be 
close to 115 GeV  (from LEP lower bound) when $f \simeq 2$ TeV we 
must have $m^2_h/f^2\simeq 3 \times 10^{-3}$. At the same time, the triplet 
must be a heavy state with $m_{\phi}\simeq f$. We therefore need
\bea
\mu^2_h &\simeq & 3 \times 10^{-3} f^2 \nn \\
\tilde{\lambda} & \simeq & 0.2 \nn\\
\lambda_{\phi} &\simeq & O(1) \, .
\eea

The condition $\lambda_{\phi}  \simeq O(1)$ yields
\be
\frac{3}{4} ( c_1 G^2+ 64 c_2  x_L^2)\simeq O(1) \, . \label{trip}
\ee
 On the other hand, the requirement (obtained by using (\ref{trip}) in (\ref{aa}))
\bea
\tilde{\lambda} & \simeq & \frac{3}{2}\, c_1 G^2 c_g^2 ( c_1 G^2 s_g^2 + 64 c_2 x_L^2) \simeq 0.2\,,
\eea
 implies that at least one of the $c_i$ coefficients must be 
fine-tuned to small 
values. 
Hence, the requirement of small values for the Higgs mass, close 
to
the experimental bound, would reintroduce the problem of fine-tuning 
that the little Higgs wanted to alleviate.

Finally, the ratio $\mu_h^2/f^2 $ is dominated by the top sector and 
is far from being of the desired order $O(10^{-3})$.
We see by \eq{mHlognum} that the problem can be ameliorated only by 
allowing the coupling $G$ to assume  large values,
and hence a very large  fine tuning between different sectors of the 
model (gauge and top loops) 
in order to cancel the top contribution to
$\mu^2_h$.
Large values of $G^2$ would also require even smaller values for the 
$c_i$ coefficients,
in order to adequately suppress $\tilde{\lambda}$.

These are all features that are confirmed by the more complete 
numerical analysis, to which we now turn, in which all the 
logarithmically divergent contributions are
properly taken into account. As discussed in the next section, the presence of the logarithmic 
contributions to the mass of the triplet will further constrain  the 
region of the allowed coefficients.

\vskip1.5em
\section{Numerical analysis} 
\label{sec:na}

Most  electroweak precision data  analyses  and  the fine tuning estimates in the littlest Higgs model  present in   literature~\cite{precision,CD} have been done expanding the CW potential up to the fourth and second order in the Higgs and triplet field respectively. In  \cite{riotto} the full potential is studied, but the logarithmic contributions are neglected. In \cite{Casas} the full potential inclusive of the logarithms is discussed.

We study the full one-loop CW potential, with no approximations both in the Higgs and in the triplet field, in order  to perform a detailed analysis of the parameter space.
As already discussed, the CW effective potential is controlled by six parameters and coefficients $(c_1, c_2, G, {G'}, x_L, f)$ which are fixed by the six constraints provided by 
\begin{itemize}
\item the existence (vanishing of first derivatives in the $h$ and $t$ directions), 
\item the value (to be $v_W$ and $v'$ for, respectively, the $h$ and $t$ fields) and 
\item the stability ($m_h^2$ and $m_t^2$ both larger than zero) 
\end{itemize}
of the ground states in the Higgs and triplet directions. 
In addition, we can add a new coefficients $c_5$ (of the two-loop correction) in order to bring $m_h$ closer to the desired value. 

Since we study the potential numerically, we  reverse the problem and instead of solving to find the values of these parameters and coefficients we generate possible sets of their values and check what  $m_h$ and $f$ (as well as the corresponding quantities for the triplet field) are thus obtained.

  We proceed in three steps by imposing the constraints which the potential must satisfy. As we shall see, these constraints greatly  reduce the allowed values of the coefficients $c_1$, $c_2$ and $c_5$. 

\begin{figure}[h]
\begin{center}
\includegraphics[width=4in]{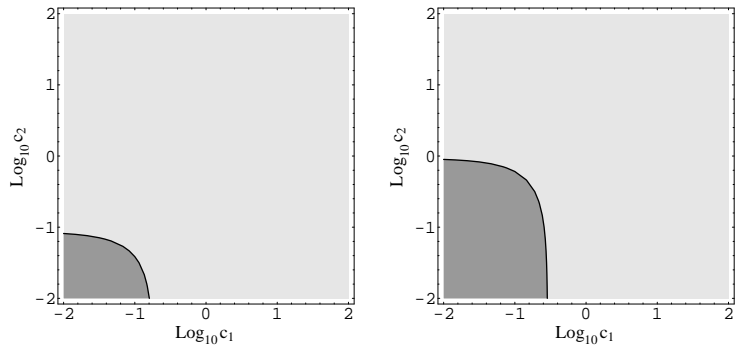}
\caption{\small Possible values (on a logarithmic scale) of the coefficients $c_1$ and $c_2$. The two figures correspond to $G'=0.72, x_L=0.56$ and, respectively, two choices of $G=3$ and $G=8$.  Each point in the light-gray region is a possible  potential with a maximum at $h/f = \pi/2$, which means  a possible minimum around $h=v_W/\sqrt{2}$. The darker region,  where both $c_i$ are small, corresponds to potentials with a minimum in $h/f=\pi/2$ which  are not allowed. \label{fig1}}
\end{center}
\end{figure}

\subsection{First step: making $v_W$ (and $v'$) the ground state}

A first constraint arises from the requirement of having the correct electroweak ground state for both the Higgs boson and the triplet fields. Here correct means for small values of the fields as opposed to larger values around $\pi f/2$. This is most easily implemented by studying
the properties of the potential along one of its direction, for instance  at large values of Higgs field $h$.
The complete potential at one-loop $V_1 [c_i,G,{G'},x_L, h/f, t/f, f]$ is a periodic function in $h/f$ in the plane where the triplet $t=0$. In order to have the ground state at the electroweak  vacuum $v_W$ around the origin,  $V_1$ must be positive for  $h/f=\pi/2$. This condition is sufficient to guarantee the existence of the correct ground state because  the complete potential for the Higgs field $h$ is given by (defining $V_1[0]=0$)
\be
V_1[h/f] = A \sin^2 h/f + B \sin^4 h/f + C \sin^6 h/f + D \sin^8 h/f
\ee
with $A,B,C$ and $D$ complicated functions of the coefficients and parameters and  such as the first derivative of the potential with respect to $h$ does not change sign between zero and $\pi/2$ when $V_1[h/f =\pi/2]<0$. Another way to understand the same feature is that if $V_1[h/f=\pi/2]$ is not positive, the mass squared of either $h$ or $t$ is negative and the electroweak ground state is unstable.

\begin{figure}[h]
\begin{center}
\includegraphics[width=4in]{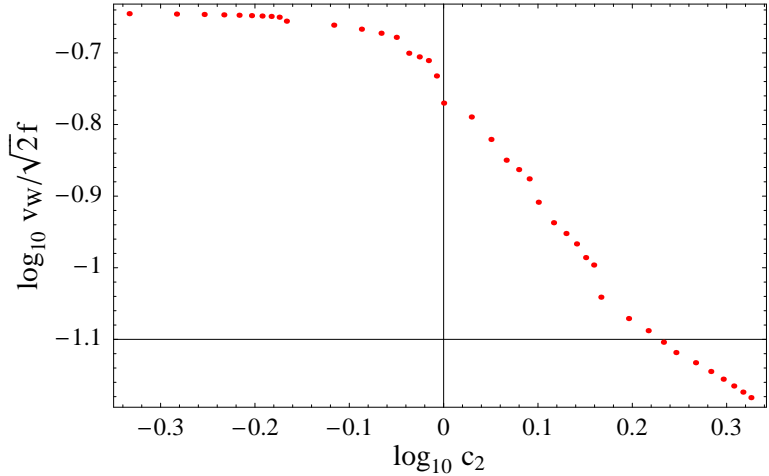}
\caption{\small Dependence (on a logarithmic scale) of the minimum on one of the coefficients after having fixed the other ($c_1=1$) and all parameters ($G=3$, $G'=0.75$ and $x_L=0.56$).  The physical region, where $f \simeq 2$ TeV, corresponds to the line $\log v_W/\sqrt{2} f = -1.1$ (lower right hand corner in the figure). \label{fig5}}
\end{center}
\end{figure}

This requirement makes possible to fix a region 
\be
V_1[c_i,G,{G'},x_L, h/f=\pi/2,t=0, f]>0
\ee
 of allowed values in the six-dimensional parameter space $(c_1,c_2,G,{G'},x_L,f)$ (no two-loop contributions are for the moment included and therefore there is no parameter $c_5$). The potential $V_1$ is given by
\bea
\frac{1}{f^4} V_1[c_i,G,{G'},x_L, h/f=\pi/2,t=0] & = & \frac{3}{16} c_1 \left( G^2 + G'^2 \right) + 12 c_2 x_L^2 + \left( \alpha - \beta\,  G^2 \right)  \log \frac{G^2}{64 \pi^2} \nn \\
&+& \left( \gamma - \delta \, G'^2  + \frac{15}{4096 \pi^2} G'^4 \right) \log \frac{G'^2}{64 \pi^2} \nn \\
&+ & \frac{3 x_L^4}{4 \pi^2 (4 x_L^2 -1)} \left( 1 + 4 x_L^2 \right) \log \frac{x_L^4}{2 \pi^2 ( 4 x_L^2 -1)} \label{Y} \, ,
\eea
where $\alpha = 6.3 \times 10^{-4}, \beta= 1.8 \times 10^{-3}, \gamma=2.0 \times 10^{-5}, \delta= 2.3 \times 10^{-4}$.
The numerical coefficients in \eq{Y} are obtained by giving their experimental values to the gauge and Yukawa couplings of the standard model.  

Fig.~\ref{fig1} shows the  values of $c_1$ and $c_2$ which satisfy the condition  above for two choices of the gauge coupling $G$ (the dependence on $G'$  is weaker). A similar plot could be shown by varying $x_L$.
In general, for given  $G,{G'},x_L$,  this condition forbids the configurations with both $c_1$ and $c_2$ of $O(10^{-2})$ and it is even more restrictive for larger values of the gauge coupling $G$ (see plot on the right side of Fig.~\ref{fig1}). 

Therefore, the very requirement of having the electroweak vacuum as the ground state of the littlest Higgs model is far from obvious for arbitrary coefficients $c_i$. As we shall see, this is important for fits to the electroweak data.  

We can  plot this ground state as a function of one of these coefficients after the other one---and all the other parameters---have been fixed to some values.    As shown in Fig.~\ref{fig5}, in the physical region, where $f \simeq 2$ TeV---which corresponds to the line $\log_{10} v_W/\sqrt{2} f = -1.1$---we obtain the desired ratio $v_W/f \sim 1/10$ for $c_1=1$ and $c_2 \simeq$ 1.2 and, therefore, with a natural choice of the coefficients.
In addition, we would also like to find $v_W\ll f$  for a large range of values of these coefficients, that is, the logarithmic derivative should not be too large:
\be
 \frac{d \log (v_W/\sqrt{2} f) }{d \log c_i}  < 10 \, .
\ee
The result in Fig.~\ref{fig5} is a variation that is  close to 1 for most values of $c_2$. In this respect,  the model is therefore working well and it stabilizes the electroweak symmetry breaking scale to be a factor of one hundred less than the cutoff of the theory.

\subsection{Second step: possible values of $m_h$ and $f$ in the one-loop CW potential}

\begin{figure}[h]
\begin{center}
\includegraphics[width=4in]{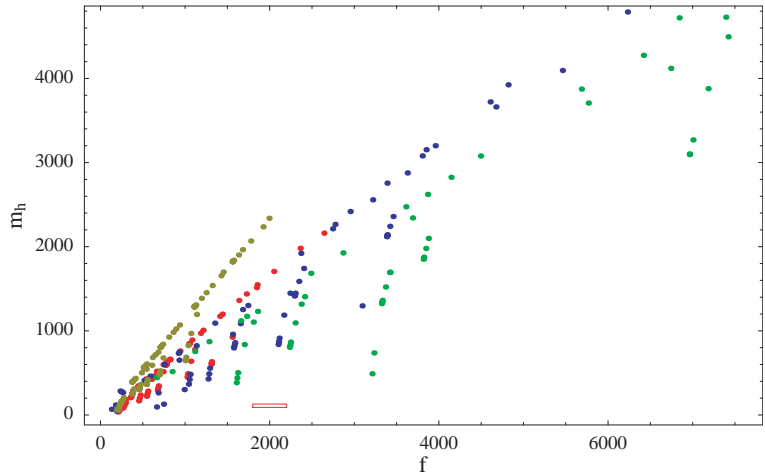}
\caption{\small $m_h$ vs.\ $f$ for $c_{1,2}$ between 0.01 and 100 Each point represent a choice of $c_1$ and $c_2$. Four different values of  $G$, $G'$ and $x_L$ ($G=1.3, 3, 8, 10$, $G'=0.72, 0.75, 2, 4$ and $x_L=0.52, 0.56, 1.2$) are shown in different colors. No 2-loop contribution is included.  The little red box (rather squeezed by the axis scales) indicates the preferred values $f=2000\pm 200$ GeV and $m_h = 110\pm20$ GeV. \label{fig2}}
\end{center}
\end{figure}

\begin{figure}[h]
\begin{center}
\includegraphics[width=4in]{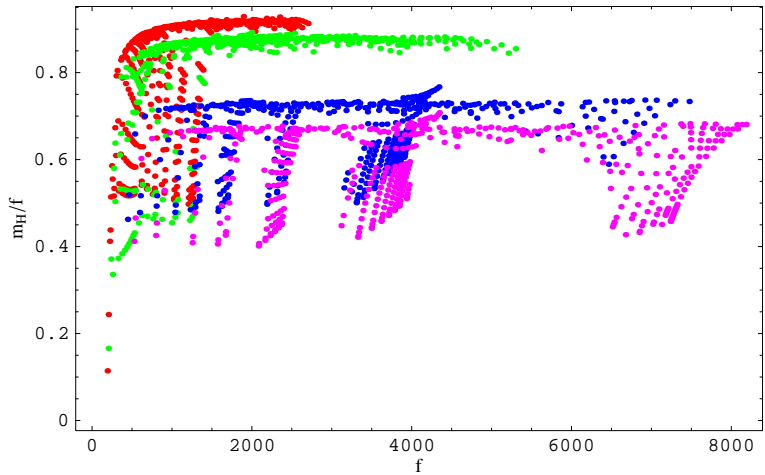}
\caption{\small $m_h/f$ vs.\ $f$ for $c_{1,2}$ between 0.01 and 100 Each point represents a choice of $c_1$ and $c_2$ with $c_1$ increasing from the bottom to the top and $c_2$ from left to right. Holes in the dots distributions are an artifact of the numerical simulation mash. Four different values of  $G=1.3, 3, 8, 12$ (at fixed $x_L=0.55$ and $G'=0.72$) are shown in different colors with smaller values toward the bottom of the figure. No 2-loop contribution is included. No choice of values of these coefficients gives a light $m_h$ and $f$ around 2 TeV at the same time.  \label{fig2a}}
\end{center}
\end{figure}
\begin{figure}[h]
\begin{center}
\includegraphics[width=4in]{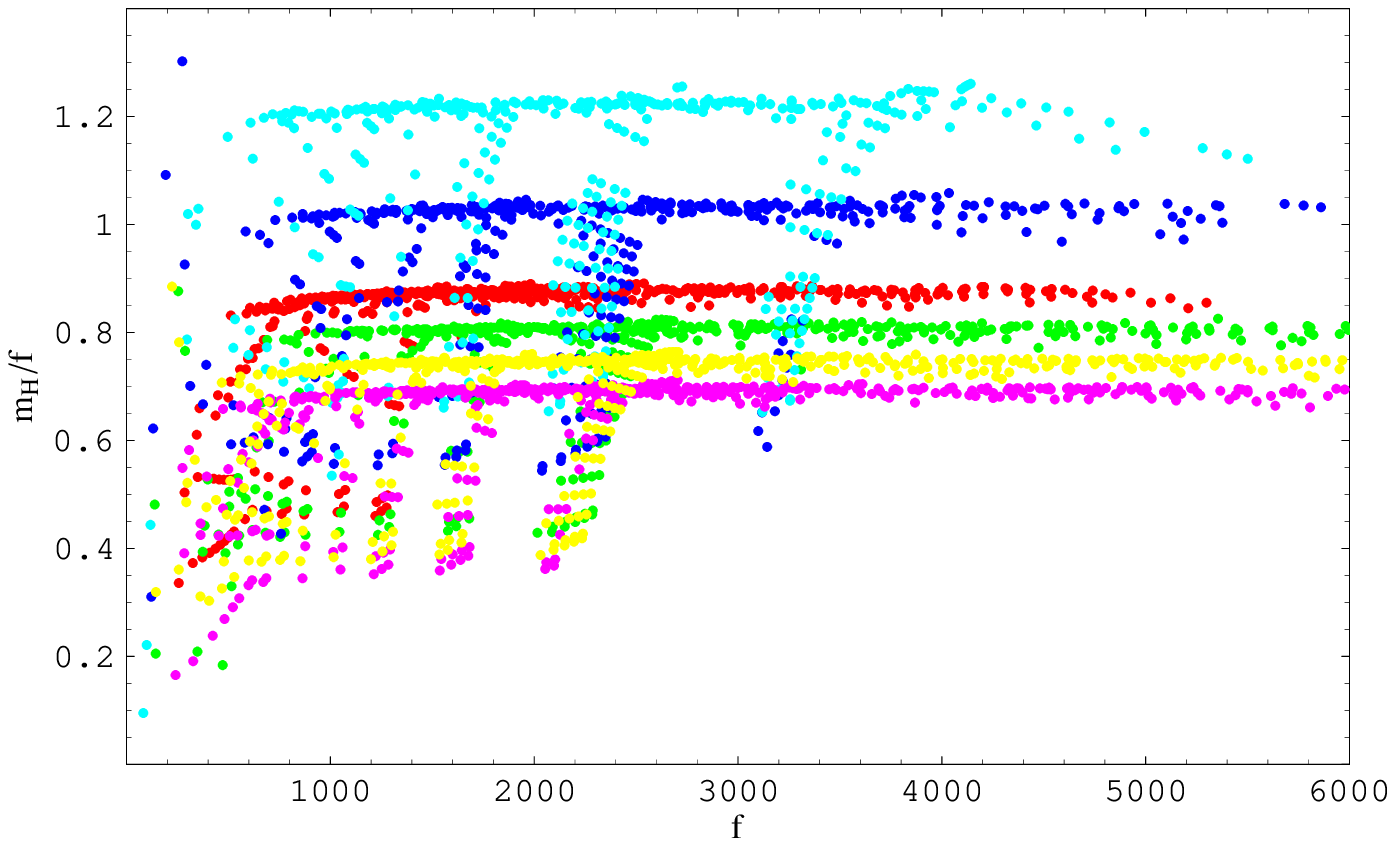}
\caption{\small Same as Fig.~\ref{fig2a}. Four different values of  $x_L=0.55, 0.71, 0.91, 1.05, 1.55, 2.05$ (at fixed $G=3$ and $G'=0.72$) are shown in different colors with the largest value of $x_L$ on top, smallest values corresponding to $x_L=0.71$. No 2-loop contribution is included. No choice of values of these coefficients gives a light $m_h$ and $f$ around 2 TeV at the same time.  \label{fig2b}}
\end{center}
\end{figure}

In the second step of our study---given the set of  parameters $(c_1,c_2,G,{G'},x_L)$ for which the scalar potential has the right behavior at large $h$ and therefore $h=v_W/\sqrt{2}$ is its ground state---we look (see Fig.~\ref{fig2}) at the   possible values of $m_h$ and $f$ for a large range of parameters and coefficients.   We take $c_i$ between 0.01 and 100, and consider four different values of $G$, $G'$ and $x_L$ to show the dependence on the gauge and Yukawa parameters. Values of $c_i$ not allowed (see Fig.~\ref{fig1}) are automatically excluded.

No choice of values gives a light mass for the Higgs boson if $f$ is larger than 1 TeV.  Roughly speaking, the mass of the Higgs boson is a linear function of the scale $f$ as we  vary $c_1$ and $c_2$. The bigger the gauge coupling $G$ (or the Yukawa $x_L$), the slower the raising of $m_h$ with $f$. Notice, however, that by increasing the value of $G$ we increase the difference in the values of the couplings $g_1$ and $g_2$ of the original gauge groups, and, for instance, at $G=10$ we find $g_1\simeq 10$ and $g_2\simeq 0.65$. The same features are also shown in Figs.~\ref{fig2a} and \ref{fig2b}, where the ratio $m_h/f$ is plotted against $f$ for different choices of $G$ and $x_L$. The natural values all lay on line at values of $m_h$ of the same order as $f$ and even stretching the parameters does not bring the ratio $m_h/f$ near the desired values (for instance, 0.1 for $f\simeq 2$ TeV). 
The dependence on $G'$ is instead rather weak. 

Even for very small $c_i$'s, the logarithmic contributions make  $m_h$  of the order of $f$ so that if we want the mass of the Higgs boson to be small, we  find that $f$ is  small as well.
Even though
it is not surprising that $m_h$ does not come out right---after all  the  (unknown and uncomputed) two-loop contributions have been usually introduced in the literature~\cite{precision} to argue that the $\mu^2$ term in the potential \eq{mu2} is essentially a free parameter to be adjusted in order to have the desired mass for the Higgs bosons---what is worrisome is that we find that  the logarithmic terms are rather large and the coefficients of the two-loop corrections would have to be accordingly large to compensate them and fine-tuned to give a net mass one order of magnitude smaller.

\subsection{Third step: including the two-loop term}

We therefore proceed to the third and final step in our analysis and include a (quadratically divergent) two-loop contribution to the term quadratic in $h$  in the scalar potential:
\be
V'[c_5;h] = \frac{c_5 f^4}{16 \pi^2} \left( \frac{h}{f} \right)^2  \, . \label{2}
\ee
This is a somewhat \textit{ad hoc} (and minimal) choice to simulate the actual 2-loop computation which is vastly more complicated and the result of which would presumably be a series of operators similar to those we have included. Other terms proportional to $\phi^2$ or $h \phi h$ could be added (and if added would completely change the analysis) but  they would correspond to two-loop corrections to already quadratically divergent one-loop terms and go against the very idea behind the little Higgs model.

\begin{figure}[h]
\begin{center}
\includegraphics[width=7in]{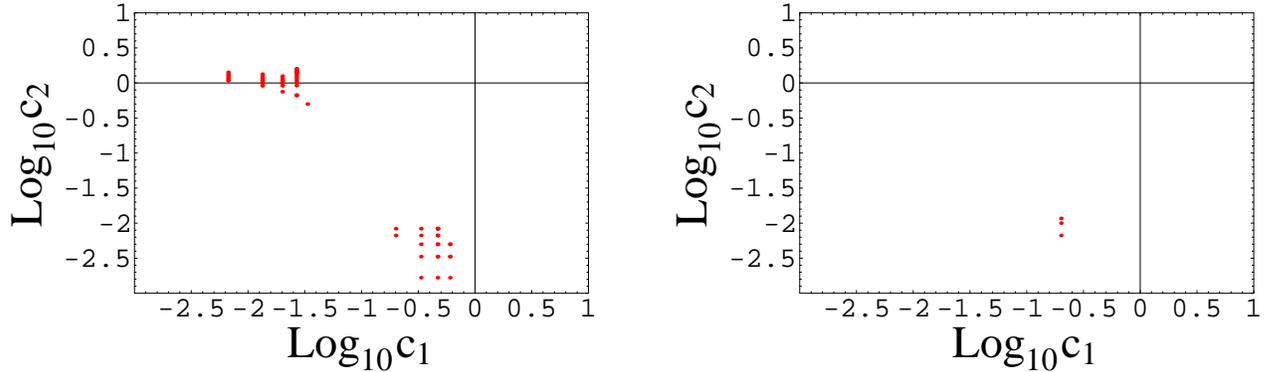}
\caption{\small  The allowed values of the coefficients $c_1$ and $c_2$ with the constrain on the Higgs boson mass ($m_h$ between 110 and 200 GeV) enforced and the two-loop quadratic divergent term included. Only very few regions in the parameter space showed in Fig.~\ref{fig1} are still allowed. On the left side: $G=3$ and $c_5\simeq 50$, on the right side: $G=8$ and $c_5 \simeq 35$. \label{fig3}}
\end{center}
\end{figure}

Having added the two-loop term (\ref{2}), it is possible to study  the behavior of the potential 
\be
 f^4 V_1 [c_i,G,{G'},x_L ,h,t,f]+ V'[c_5,h] \, . \label{pot2}
\ee
around the origin. For each choice of $(G,{G'},x_L)$, by imposing the four constrains arising from the two first derivatives  (to have a minimum and it to be at the correct value) and from the value of $m_h$ and $m_t$, the three coefficients, $c_1,c_2$ and $c_5$, are fixed. 

The study we performed shows that the new constraint of having a Higgs boson mass close to the current bound~\cite{PDG} drastically reduces the allowed region in the parameter space 
of $(c_1,c_2,G,{G'},x_L,f)$ and given $G$, ${G'}$ and $x_L$ the allowed regions are characterized by having either $c_1$ of $O(1)$ and  $c_2$ of $O(10^{-2})$ or the opposite, as shown in Fig.~\ref{fig3}. For each of these choices of coefficients $c_1$ and $c_2$, a value of $c_5$ must be chosen so as to obtain the desired mass $m_h$. This is only possible for rather large values of the coefficient $c_5$. If we are willing to allow larger Higgs boson masses (that is, $m_h > 300$ GeV), $c_5$ will turn proportionally smaller but we will still have similar severe constraints on $c_1$ and $c_2$.

\begin{figure}[h]
\begin{center}
\includegraphics[width=4in]{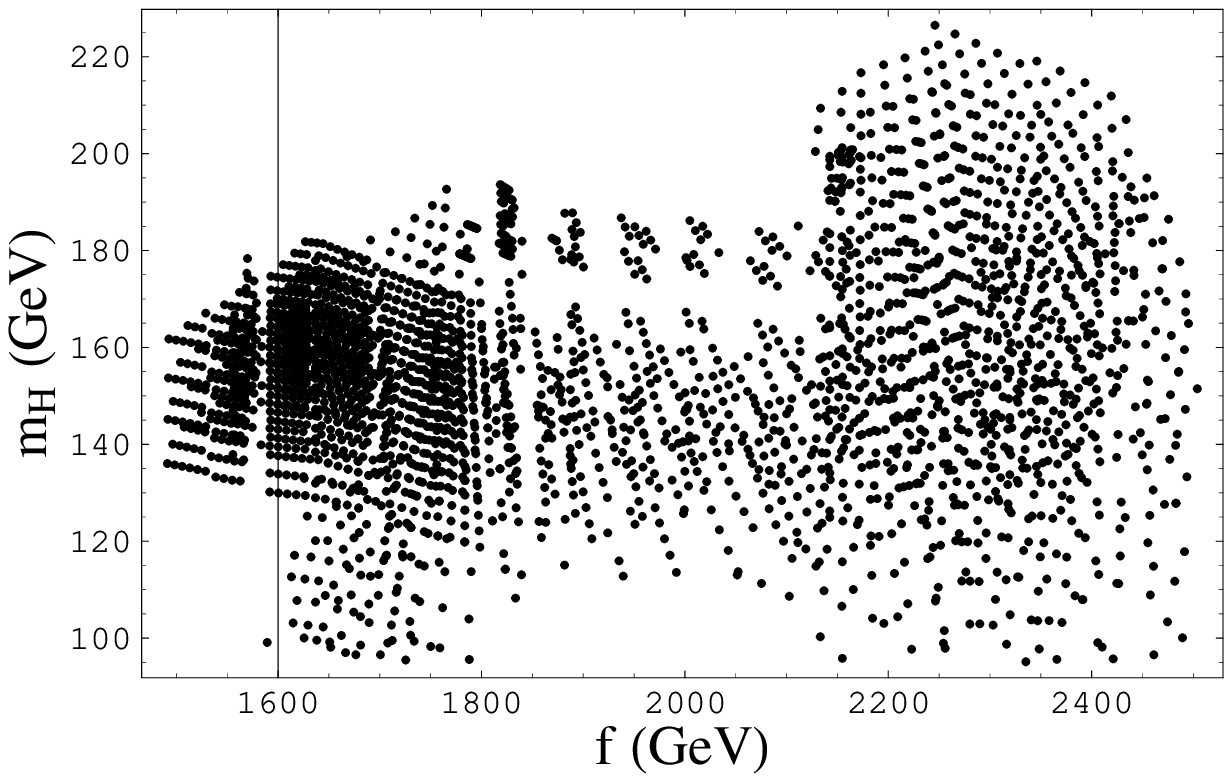}
\caption{\small$m_h$ vs.\ $f$. Each point corresponds to a choice of all coefficients and parameters in the range discussed in the text ($c_1= O(10^{-2})$, $c_2=O(1)$, $c_5\simeq 50$) and varied in discrete steps; $G=3$, $G'=0.75$ and $x_L=0.56$ are fixed. \label{fig4} }
\end{center}
\end{figure} 

\subsection{Conclusions}

Fig.~\ref{fig4} shows the possible values of $m_h$ and $f$ close to the desired values for a range of the coefficients $c_1$, $c_2$ and $c_5$ in the allowed regions.
 These values are  now possible but we pay a rather high price for it. The two main problems are that
\begin{itemize}
\item the natural case in which all the coefficients $c_i$ are $O(1)$ seems to be ruled out.  Values for $f$ and $m_h$ in the desired range are only obtained by taking $c_1$ of $O(10^{-2})$ and $c_2$ $O(1)$ or vice versa.  A coefficient of order $O(10^{-2})$ clearly goes against the very rationale of introducing the littlest Higgs model in the first place because we have to make small by hand one of the symmetry breaking terms;
\item the phenomenological two-loop term must have rather large coefficients ($c_5=$45-55). This already anticipated feature reminds us of the importance of these terms in compensating the  logarithmic contribution to the Higgs boson mass, which are therefore rather larger than one would wish and usually assume in the little-Higgs  framework.  Roughly speaking,  these logarithmic terms are $O(f)$ whereas we expected them to be of $O(m_h)$. This is unfortunate since the naturalness of a scale $f$ around 2 TeV is questionable once such a large two-loop term is included in order to bring $m_h$ around its current bound. Moreover, given the size of our example of two-loop contribution, there is no way to argue that these two-loop contributions can be neglected in any other part of the potential and the entire approach at one-loop seems to break down. A similar conclusion was reached in a recent work were the fine-tuning of the littlest Higgs is discussed~\cite{Casas}.
\end{itemize}

The analysis above shows that once the scale $f$ is required to be larger than 1 TeV, after all coefficients have been fixed, the value of the Higgs boson mass---which is linked to that of the neutral component of the triplet---cannot be made as small as desired. In particular, it is not possible to have it close to the current experimental lower bound unless some the coefficients of the quadratically divergent terms are made unrealistically small while at the same time the two-loop correction is made rather large. The necessary smallness of some of the coefficients defeats the purpose of introducing the collective breaking mechanism to make the mass terms small and the littlest Higgs model stable against one-loop radiative corrections. Moreover, the mass of the Higgs boson itself comes out in a very unnatural  way from the cancellation of terms one order of magnitude larger than its value.

This result seems to be a more serious problem for the model than that of the fine tuning required in order to be consistent with electroweak precision measurements. The problem has been so far ignored in the literature because it has been assumed that it was always possible to add to  the logarithmically divergent part of the potential the two-loop quadratically divergent contribution so as to obtain the desired Higgs boson mass. This is however only possible at the price of introducing an unreasonable large coefficient in this term and even then at the price of having at least one of the other two coefficients very small.  

On the other hand, if we let instead the model to decide what value the mass of the Higgs boson should be, we find that it comes out close to the scale $f$ and therefore, for $f$ in the 1-2 TeV range, the Higgs is accordingly heavier than expected. 
A similar result (but for different reasons) was recently obtained in a little-Higgs-like unified model of electroweak and flavor physics~\cite{flhiggs}. This result is not necessarily in contradiction with the electroweak precision measurements~\cite{Peskin} because the fit should now be redone after including the heavy Higgs boson as well as the new states introduced by the littlest Higgs model (see, however, \cite{strumia}).

\subsection{Comparison with other studies}

There are many discussions in the literature about  the littlest Higgs model and electroweak precision constraints~\cite{precision,CD}.
In all these papers, however, the values of the coefficients of the divergent terms are  assumed to be of $O(1)$ or, at most $O(0.1)$ and the logarithmic terms not included.  Moreover in \cite{precision} only the parameters relevant to the effective operators in the gauge boson sector are discussed and the coefficients of the divergent terms are assumed of the desired order and not studied. The only reference in which the scalar potential is actually constrained is \cite{CD}. In order to show that our conclusions agree with what found in this reference,  let us, following their notation, fix the coupling $g$ and $g^\prime$ in terms of the fine-structure constant $\alpha$ and the Weinberg angle, $v_W$ and $v^\prime$---the  vacuum expectation values of the isospin triplet $t$---by means of the Fermi constant and reparametrize the top Yukawa couplings $\lambda_1$ and $\lambda_2$ in terms of
\bea
x_L  & = & \frac{\lambda_1^2}{\lambda_1^2 + \lambda_2^2} \nn \\ 
\frac{m_t}{v_W} &=&  \frac{\lambda_ 1\lambda_2}{\sqrt{\lambda_1^2 + \lambda_2^2}} \left[ 1 + \frac{v^2}{2 f^2} x_L ( 1 + x_L) \right]
\eea
we are thus left with a model that, after assigning a value to $m_t$ and $m_H$, only depends on $f$, $x_L$, $s$ and $s^\prime$ (as defined in Ref.~\cite{CD}) and the counterterms $a$ and $a^\prime$ (which correspond to  $3 c_1/2$ and $6 c_2$). These two can be found for each choice of the first four parameters by solving 
\bea
\frac{a}{2} \left[ \frac{g^2}{s^2c^2} + \frac{{g^\prime}^2}{{s'}^2 {c'}^2} \right] + 8 a^\prime \lambda_1^2 & = & 2 \frac{m_H^2}{v_W^2} \frac{1}{1 -\left( 4 v^\prime f/v_W^2 \right)^2} \nn \\
- \frac{a}{4} \left[ \frac{g^2 (c^2-s^2)}{s^2c^2} + \frac{{g^\prime}^2 ({c'}^2 - {s'}^2)}{{s'}^2 {c'}^2} \right] + 4 a^\prime \lambda_1^2 & = & 2 \frac{m_H^2 v^\prime f}{v_W^4} \frac{1}{1 -\left( 4 v^\prime f/v_W^2 \right)^2}
\eea
 We thus find that in order to have, for instance, $f= 2$ TeV while $m_H=115$ GeV (and $v' = 3.54$ GeV, $x_L=0.4$, $s=0.22$ and $s'=0.66$, as discussed in \cite{CD}) we must take the coefficients $a$ and $a'$ of order $1/100$ (more precisely, $a=0.036$ and $a'=0.063$ in this case; small coefficients are found also for other allowed choices of $f$ and $v'$), a choice that clearly defeats the very rationale for introducing the littlest Higgs model in the first place. 
 
 This result is consistent with our analysis as presented in the previous section in the case in which the logarithmic contributions are neglected and the 2-loop terms included. However, as soon as the logarithmic contributions are not neglected (and we have shown that they cannot be neglected), the solution above does not exist because it would correspond to a negative value of the triplet mass and an unstable electroweak ground state. Going back to Fig.~\ref{fig1}, the solutions studied in \cite{CD} is in the region ruled out where both coefficients $c_i$ are very small.

\vskip1.5em
\section{A modified top sector} 
\label{sec:nm}

In the previous sections we have seen  that the littlest Higgs  model, given a cutoff $\Lambda= 4 \pi f$ around $10$ TeV, predicts a large Higgs mass around $500$ GeV. Introducing a 2-loop effective  quadratic term allows to bring this mass to a value smaller than $200$ GeV but the 
2-loop term coefficient must then be very large. Because the problem is largely due to the fermionic sector of the model, in this section we  discuss a possible modification of the fermion content of the model, as proposed   in \cite{nelson}, to see if it helps.  We neglect in the following the triplet
and focus on the Higgs doublet.   

We anticipate that, with these modifications, the fermion contribution to the
potential is finite, and hence no ambiguity (or freedom) is left in the choice
of the coefficients of this part of the potential, which are fixed by the choice of
(physical) masses and couplings.
Here we report only on the approximate analysis of the model,
which is confirmed by the numerical study we performed,
since the results are not significantly different from the 
model discussed previously.

\subsection{The model}

The lagrangian of the model we consider differs from that of the littlest Higgs model  only by the fermionic contributions $\mathcal{L}_\psi$. The fermionic content is given by an  electroweak doublet $Q_L=(t_0,b_0)_L$, an electroweak singlet  $t^c_{g_{L}}$ and by two colored $SU(5)$ quintuplets, $X$ and $\bar{X}$. The Yukawa lagrangian 
is  given by an $SU(5)$ invariant term, and by two explicit breaking terms of the $SU(5)$ global symmetry, both of them preserves enough symmetry to prevent the Higgs to gain a mass.  Only the loops contributions that involves both of them can produce a mass for the Higgs. The Yukawa lagrangian is given by
\be
\label{eqf}
\mathcal{L}_Y= \sqrt{2} \lambda_1 f\, \bar{X}\,\Sigma\, X+ \sqrt{2}\lambda_2 f\, (a^c_{1_{L}} b_{0_{L}} +a^c_{2_{L}} t_{0_{L}})  +\sqrt{2} \lambda_3 f\, t^c_{g_{L}} t_{d_{L}}  \,,
\ee
where 
\bea
X=\left( \begin{array}{c}p_{1_{L}}\\
p_{2_{L}}\\t_{d_{L}}\\r_{1_{L}} \\r_{2_{L}} 
\end{array}\right)& & \bar{X}=\left( \begin{array}{c} a^c_{1_{L}}\\
a^c_{2_{L}}\\t^c_{t_{L}}\\b^c_{1_{L}} \\b^c_{2_{L}} 
\end{array}\right)\,.
\eea

By \eq{eqf} we obtain the fermion mass matrix given by
\bea
\label{mf}
M_{f_{RL}}&=& f\, \left(\begin{array}{cccc}
 -\sqrt{2} \lambda_1 \,\sin^2\frac{h}{f}& i\,\lambda_1\, \sin \frac{2h}{f}   &\sqrt{2} \,\lambda_2&\sqrt{2}\, \lambda_1\, \cos^2\frac{h}{f} \\
\sqrt{2}\, \lambda_1 \,\cos^2\frac{h}{f}& i\,\lambda_1 \,\sin \frac{2h}{f} &0 & -\sqrt{2} \,\lambda_1 \,\sin^2\frac{h}{f} \\0& \sqrt{2}\, \lambda_3& 0&0 \\
i\,\lambda_1 \,\sin \frac{2h}{f}&i\,\sqrt{2}\,\lambda_1 \,\cos \frac{2 h}{f}&0& i\,\lambda_1 \,\sin \frac{2h}{f}
\end{array} \right) \,,
\eea
where in \eq{mf} we have put $t=0$.
By \eq{mf} we see that 
\bea
\label{trace}
\Tr M^\dag_{f_{RL}}M_{f_{RL}}&=& 2 ( L_1^2+ L_2^2 +  \lambda_1^2)\, f^2 \nn \\
\Tr (M^\dag_{f_{RL}}M_{f_{RL}})^2&=& 4 ( L_1^4+ L_2^4 +  \lambda_1^4)\, f^4 \,,
\eea
where 
\bea
\label{LL}
L_1^2&=&\lambda_1^2+\lambda_2^2 \nn \\
L_2^2&=&\lambda_1^2+\lambda_3^2 \,. 
\eea
Eq.~(\ref{trace}) indicates that there are no one-loop fermionic divergent contributions to the mass of the Higgs. The only  one-loop  fermionic contributions are finite and therefore calculable. From now on we take $L_1=L_2$.

One of the eigenvalues does not depend on $h$, and is given by:
\be
m^2_3= 2 \lambda_1^2 f^2 \,. 
\ee
The lightest mass is to be interpreted as that of the standard top quark, with mass approximated by
\be
\label{masst}
m^2_t =  \lambda_t^2 f^2 \sin^2 \frac{h}{f} +\left(-\lambda_t^2+\frac{\lambda_t^4}{L_1^2}\right)\sin^4 \frac{h}{f} \,,
\ee
where 
\bea
\label{ltop}
\lambda_t &=& 2\frac{\lambda_1 \lambda_2 \lambda_3}{\sqrt{\lambda_1^2+\lambda_2^2}\sqrt{\lambda_1^2+\lambda_3^2}} \, .
\eea
It gives a negligible contribution to the effective potential.
The two relevant eigenvalues can be expanded in powers of $\sin h/f$ obtaining:
\bea
\label{casouguali}
m^2_1/f^2& =& 2 L_1^2 + \sqrt{2} L_1 \lambda_t \sin\frac{h}{f}-\frac{\lambda_t^2}{2}  \sin^2\frac{h}{f}- \left(\frac{L_1\lambda_t}{\sqrt{2}}-\frac{5\lambda_t^3}{8\sqrt{2}L_1}\right) \sin^3\frac{h}{f}\nn \\
&&+\frac{1}{2}\left(\lambda_t^2-\frac{\lambda_t^4}{L_1^2}\right) \sin^4\frac{h}{f}\,  + O(\sin^5\frac{h}{f})\, ,\nn \\
m^2_2/f^2& =& 2 L_1^2 - \sqrt{2} L_1 \lambda_t \sin\frac{h}{f}-  \frac{\lambda_t^2}{2}\sin^2\frac{h}{f} +\left(\frac{L_1\lambda_t}{\sqrt{2}}-\frac{5\lambda_t^3}{8\sqrt{2}L_1}\right) \sin^3\frac{h}{f}\nn\\
&&+\frac{1}{2}\left(\lambda_t^2-\frac{\lambda_t^4}{L_1^2}\right) \sin^4\frac{h}{f}\,+  O(\sin^5\frac{h}{f}) \,.
\eea

The fermionic contribution to the potential   for the Higgs field  obtained when $L_1 = L_2$ (which is the most favorable case) is therefore 
\be
\label{potf}
\frac{V_{tn}}{f^4}\,=\,-\frac{3\lambda_t^2 L_1^2}{4\pi^2}\,\sin^2\frac{h}{f}\,+\,
\frac{\lambda_t^4}{16\pi^2}\,\left(-4+\frac{12L_1^2}{\lambda_t^2}-3\ln\frac{\lambda_t^2}{2L_1^2}\right)\,\sin^4\frac{h}{f}\,.
\ee

\subsection{Approximate analysis}

The bosonic sector of the model has not been modified, hence we 
can write (see \eq{V0ap}) the (approximate) expressions:
\bea
\label{mHlogvar}
\frac{\mu_h^2}{f^2} &=& - \frac{9}{256 \pi^2} g^2 G^2 
\log \frac{G^2}{64 \pi^2}  -\frac{3\lambda_t^2 L_1^2}{4\pi^2}\,,\nn \\
\lambda_4&=&\frac{3 c_1G^2}{16}\,+\, \frac{\lambda_t^2 L_1^2}{4\pi^2}\,+\,
\frac{\lambda_t^4}{16\pi^2}\,\left(-4+\frac{12L_1^2}{\lambda_t^2}-3\ln\frac{\lambda_t^2}{2L_1^2}\right)\,.
\eea
Choosing $L_1\sim \sqrt{2}$ (that is, a value close to the smallest possible after $\lambda_t=1$), one obtains (taking the bosonic part from \eq{mHlognum})
\bea
\frac{\mu_h^2}{f^2} &\simeq&0.01 G^2 g^2  -0.15\,,\nn \\
 \lambda_4&\simeq&\frac{3 c_1 G^2}{16}+0.2\,.
\eea

The finite contributions to $\mu_h^2$
in this variation of the model 
are comparable in size to the original logarithmically divergent ones.
Further, the quartic coupling is now dominated by the gauge boson sector,
since the top sector gives only a small contribution.
From this, comparing with the original littlest Higgs model, we conclude that
there is no substantial improvement: the cancellation of logarithmic 
divergences is not enough to reduce the large top 
contribution to the $\sin^2h/f$ term in the potential. For this reason we leave out a more general numerical analysis of this modified model. 

\vskip1.5em
\section{The littlest Higgs model at finite temperature} 
\label{sec:ft}

We now turn to the study of the littlest Higgs model at finite temperature to determine the existence and nature of its phase transitions. We do it by assuming that the Higgs mass is in the light, notwithstanding our argument against this choice, because it is the scenario more often discussed in the literature and also because the phase transition can only become weaker for a heavier Higgs mass.

The finite temperature effective potential is given by
\be
V[c_i,\Sigma,T]= V_1[c_i,\Sigma]+  V_T[c_i,\Sigma]
\ee
 where $ V_1[c_i,\Sigma]$ is the potential of \eq{V0} and $V_T[c_i,\Sigma]$
is the temperature dependent contribution
\be
V_T[c_i,\Sigma] = \mp g_f \frac{T^4}{2 \pi^2} \int^\infty_0 dx\, x^2 \ln \left[ 1 \pm \exp -\sqrt{x^2 + M^2(\Sigma)/T^2} \right] \, , \label{potTtrue}
\ee
where the sign of the exponential term depends on the statistics of the particles and $g_f$ is the number of degrees of freedom. In the limit $m_i/T \ll 1$ , with $m_i$ the mass of a generic boson or fermion, \eq{potTtrue} simplifies and we obtain
\bea
\label{VT}
 V_T[c_i,\Sigma]&\simeq &  \frac{T^2}{24} \left[\Tr{M^2_B(\Sigma)}+\frac{1}{2}\Tr{M^2_F(\Sigma)}\right] -\frac{T}{12 \pi}\Tr{M^3_B(\Sigma)} \nn\\
&+&\frac{1}{64 \pi}\Big[\Tr{M^4_B}(\Sigma)\big(\log\frac{c_B T^2}{\Lambda^2} \big)-\Tr{M^4_F}(\Sigma)\big(\log\frac{c_F T^2}{\Lambda^2}\big)\Big] \,,
\eea
where $B,F$ denote, respectively, the bosonic and fermionic degrees of freedom.   

Because the potential is very different, at least at large $h$, with respect to that of the standard model, one may wonder whether the electroweak phase transition is any stronger for  values of the Higgs boson mass close to the current bounds than in the standard model. This is an important problem in the study of baryogenesis.

\begin{figure}[h]
\begin{center}
\includegraphics[width=4in]{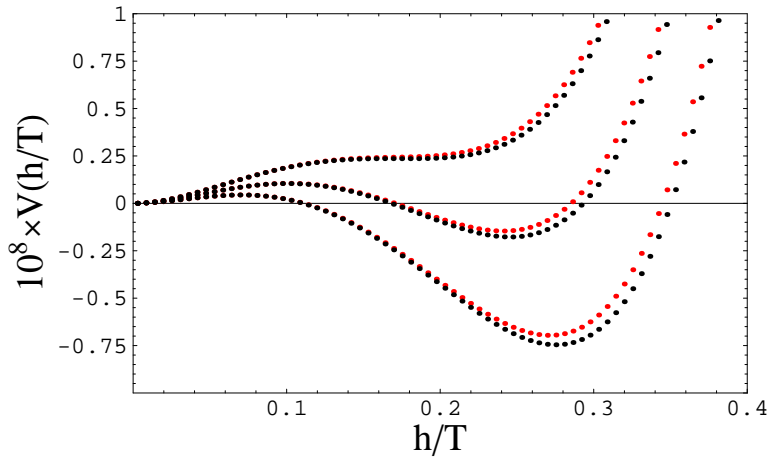}
\caption{\small  A comparison  of the potential $V[h]$ in the standard model and in the littlest Higgs model ($G=1.3$, $G'=0.72$, and $x_L=0.56$) at three different $T$ close to $T_c$. The coefficients $c_1$, $c_2$ and $c_5$ of the CW potential are those yielding a small Higgs mass. Red dots represent the standard model, black dots the littlest Higgs model behavior.  The transition is weakly of the first order for $m_h=120$ GeV for both models.  \label{fig7}}
\end{center}
\end{figure}

We study the potential with the restrictions on the coefficients and parameters we have discussed so far (that is, the coefficients $c_i$ in the range of Fig.~\ref{fig4}). Fig.~\ref{fig7} shows the  potential of the littlest Higgs model and compares it to that of the standard model at three different $T$ close to  $T=T_c$, where $T_c$ is the temperature of the electroweak phase transition.  A small improvement is present in going from the standard model to the littlest Higgs but it is not significant. The crucial term linear in $T$ is not large enough to strengthen the transition and, as in the standard model, only for  small values of the Higgs boson mass the phase transition can be strong enough to sustain electroweak baryogenesis.

Models based on pseudo-Goldstone bosons may present an interesting phenomenon of symmetry non-restoration at high $T$ (see~\cite{Kolb}, and more recently \cite{riotto} in the little-Higgs context). The littlest Higgs model is case in point. Because the potential is a periodic function of the pseudo-Goldstone fields, and of the Higgs field $h$ in particular, as the temperature increases, the maximum in the potential at $h/f=\pi/2$ turns into  a minimum with an energy lower than in $h=0$. Accordingly, the symmetric ground state in zero becomes unstable and there is no restoration of the symmetry at higher temperatures.

\begin{figure}[h]
\begin{center}
\includegraphics[width=4in]{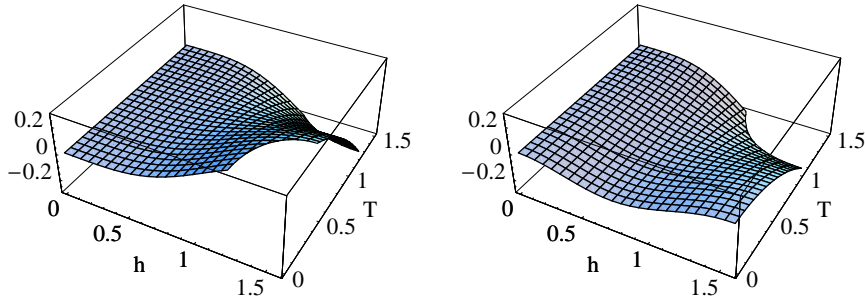}
\caption{\small   Comparison between the potential as a function of $h$ and $T$ in the littlest Higgs model without and with the logarithmically divergent terms ($G=3$, $G'=0.75$, and $x_L=0.56$). The value of $T_c$ where $V[h_{min}, T_c]=0$ changes by 20\% (from $T_c \simeq f$ to $0.7 f$) after including the logarithmic terms.\label{fig8}}
\end{center}
\end{figure}

When we study the  high temperature behavior, we can approximate \eq{VT} by
\be
V_T[c_i,\Sigma] \simeq \frac{T^2}{24} \Big(\Tr{M^2_B(\Sigma)}+\frac{1}{2}\Tr{M^2_F(\Sigma)}\Big)
\ee
Keeping into account for the generic $M^2_{B,F}(\Sigma)$ only the 1-loop quadratically divergent contributions, we have:
\bea
\sum_{d.o.f.}\Tr{M^2_V}&=&  \frac{9}{4}G^2+ \frac{3 G'^2}{20}+ \frac{3}{16}  ( G^2+G'^2)\sin^4{h/f}   \nn\\
\sum_{d.o.f.}\Tr{M^2_{S,PS}} &=& \frac{3}{2} \left[ c_1(G^2+G'^2)+ 64 c_2 x_L^2 \right] (1-2 \sin^4{h/f}) \nn\\
\sum_{d.o.f.}\Tr{M^2_{SC}} &=& \frac{3}{2} \left[ c_1(G^2+G'^2)+ 64 c_2 x_L^2 \right]  (1- \sin^4{h/f} ) \nn\\
\sum_{d.o.f.}\Tr{M^2_{DC}} &=& \frac{3}{2} \left[ c_1(G^2+G'^2)+ 64 c_2 x_L^2 \right] (1- \sin^4{h/f}\big) \nn\\
\sum_{d.o.f.}\Tr{M^2_{F}} &=& \frac{8 x_L^4}{4 x_L^2-\lambda_t^2}-x_L^2\sin^4{h/f} \, ,
\eea 
 where $V$,  $S,PS$, $SC$, $DC$ denote respectively the gauge bosons contributions, the scalar and pseudoscalar contributions and the single and double charged ones.
 
 We therefore have  the  potential 
 \bea
 V[c_i, h, T] & \simeq & \left\{ \frac{3}{16} c_1 (G^2 + G'^2) + 12  c_2 x_L^2 + T^2 \left[- \frac{3}{16} c_1 (G^2 + G'^2) \right. \right. \nn  \\ 
 & -& \left. \left.  16  c_2 x_L^2 + \frac{1}{192} ( 3 G^2 + 3 G'^2 - 32 x_L^2) \right] \right\} \sin^4 h/f \, , \label{potT}
 \eea
  which is a good approximation at high temperature. 

 The potential in (\ref{potT})  clearly depends on the coefficients $c_i$ and the very presence or not of a phase transition depends on  their values. For arbitrary choices the phase transition can be anywhere and even not exist at all. However, we find that for values of these coefficients in the allowed range---where not both coefficients are small---we identified in the previous sections, the phase transition is always present and for values of the temperature $T< \Lambda/\pi \simeq 4 f$ for which we trust the potential.

 The potential (\ref{potT})  only includes the quadratically divergent terms. As discussed in the previous sections, we must add to it the logarithmically divergent terms as well in order to obtain a reliable result.
 
 Fig.~\ref{fig8} shows the behavior of the  potential for different $T$ and  $h$ and compares the case without the logarithmic terms  with that in which all terms in the potential are retained. The phase transition is always present but the value of $T_c$ is moved by a substantial amount (20\%) so that it is necessary to keep the full potential if we want to discuss the temperature dependence of the potential. In particular, $T_c$ tends to be smaller than $f$ after including the logarithmic terms and this means that the potential in \eq{potT} is not correct because some of the heavy states  may now live above the critical temperature obtained within the approximated potential,  and the exact form based on \eq{potTtrue} should be used instead.
  
\acknowledgments

 MP acknowledges SISSA and INFN, Trieste, for hospitality 
during the completion of this research. 
This work is
partially supported by  the European TMR Networks HPRN-CT-2000-00148
and HPRN-CT-2000-00152. The work of MP is supported in part by the
US Department of Energy under contract DE-FG02-92ER-40704.


 \end{document}